\documentclass[10pt,conference]{IEEEtran}

\ifx\pdfoutput\undefined
\usepackage[hypertex]{hyperref}
\else
\usepackage[pdftex,hypertexnames=false,hyperfigures=false,pdfview=FitH,pdfstartview=FitH]{hyperref}
\fi

\usepackage{amsmath,amssymb}
\usepackage{cite}

\def\ol#1{\overline{#1}}
\def\C{{\mathbb C}}
\def\F{{\mathbb F}}
\def\Z{{\mathbb Z}}
\def\entspricht{\mathrel{\hat{=}}}
\def\dist{\mathop{\rm dist}}
\def\wgt{\mathop{\rm wgt}}
\def\bra#1{\langle#1|}
\def\ket#1{|#1\rangle}

\newtheorem{theorem}{Theorem}
\newtheorem{definition}[theorem]{Definition}
\newtheorem{lemma}[theorem]{Lemma}

\begin{document}

\title{Quantum Goethals-Preparata Codes}


\author{\IEEEauthorblockN{Markus Grassl}
\IEEEauthorblockA{
Institute for Quantum Optics and Quantum Information\\
Austrian Academy of Sciences\\
Technikerstra{\ss}e 21a, 6020 Innsbruck, Austria\\
Email: markus.grassl@oeaw.ac.at}
\and
\IEEEauthorblockN{Martin R\"otteler}
\IEEEauthorblockA{
NEC Laboratories America, Inc.\\
4 Independence Way, Suite 200\\
Princeton, NJ 08540, USA\\
Email: mroetteler@nec-labs.com
}
}

\maketitle

\begin{abstract}
We present a family of non-additive quantum codes based on Goethals
and Preparata codes with parameters $((2^m,2^{2^m-5m+1},8))$.  The
dimension of these codes is eight times higher than the dimension of
the best known additive quantum codes of equal length and minimum
distance.
\end{abstract}


\section{Introduction}
Most of the known quantum error-correcting codes (QECCs) are based on
the so-called stabilizer formalism which relates quantum codes to
certain additive codes over $GF(4)$ (see, e.\,g.,
\cite{CRSS98,Got96}).  It is known that non-additive QECCs can have a
higher dimension compared to additive QECCs with the same length and
minimum distance\cite{CSSZ07,RHSS97,YCLO07,YCO07}.  All these examples
of non-additive QECCs are examples of so-called codeword stabilized
quantum codes which are obtained as the complex span of some so-called
stabilizer states, which correspond to self-dual additive codes.  In
\cite{GrRo07} we have extended the framework of stabilizer codes to
the union of stabilizer codes (see \cite{GrBe97}). This allows to
construct non-additive codes from any stabilizer code. In general,
these non-additive QECCs correspond to non-additive codes over $GF(4)$
which can be decomposed into cosets of an additive code which contains
its dual.  Using a construction similar to that of so-called CSS codes
(see \cite{CaSh96,Ste96:simple}), families of non-additive quantum
codes based on the binary Goethals and Preparata codes were derived in
\cite{GrRo07}.  Here we present a new family of non-additives quantum
codes which have a dimension that is eight times higher than the
dimension of the best known additive quantum codes.

\section{Union Stabilizer Codes}
\subsection{Stabilizer codes}
We start with a brief review of the stabilizer formalism for quantum
error-correcting codes and the connection to additive codes over
$GF(4)$ (see, e.\,g., \cite{CRSS98,Got96}). A stabilizer code encoding
$k$ qubits into $n$ qubits having minimum distance $d$, denoted by
${\cal C}=[[n,k,d]]$, is a subspace of dimension $2^k$ of the complex
Hilbert space $(\C^2)^{\otimes n}$ of dimension $2^n$.  The code is
the joint eigenspace of a set of $n-k$ commuting operators
$S_1,\ldots,S_{n-k}$ which are tensor products of the Pauli matrices
\[
\sigma_x=\left(\begin{matrix}0&1\\1&0\end{matrix}\right),\quad
\sigma_y=\left(\begin{matrix}0&-i\\i&0\end{matrix}\right),\quad
\sigma_z=\left(\begin{matrix}1&0\\0&-1\end{matrix}\right),
\]
or identity.  The operators $S_i$ generate an Abelian group ${\cal S}$
with $2^{n-k}$ elements, called the \emph{stabilizer} of the code.  It
is a subgroup of the $n$-qubit Pauli group ${\cal P}_n$ which itself
is generated by the tensor product of $n$ Pauli matrices and identity.
We further require that ${\cal S}$ does not contain any non-trivial
multiple of identity.  The {\em normalizer} of ${\cal S}$ in ${\cal
  P}_n$, denoted by ${\cal N}$, acts on the code ${\cal C}=[[n,k,d]]$.
It is possible to identify $2k$ logical operators
$\ol{X}_1,\ldots,\ol{X}_k$ and $\ol{Z}_1,\ldots,\ol{Z}_k$ such that
these operators commute with any element in the stabilizer ${\cal S}$,
and such that together with ${\cal S}$ they generate the normalizer
${\cal N}$ of the code.  The operators $\ol{X}_i$ mutually commute,
and so do the operators $\ol{Z}_j$. The operator $\ol{X}_i$
anti-commutes with the operator $\ol{Z}_j$ if $i=j$ and otherwise
commutes with it.

It has been shown that the $n$-qubit Pauli group corresponds to a
symplectic geometry, and that one can reduce the problem of
constructing stabilizer codes to finding additive codes over $GF(4)$
that are self-orthogonal with respect to a symplectic inner product
\cite{CRSS96,CRSS98}. Up to a scalar multiple, the elements of ${\cal
  P}_1$ can be expressed as $\sigma_x^a\sigma_z^b$ where
$(a,b)\in\F_2^2$ is a binary vector. Choosing the basis $\{1,\omega\}$
of $GF(4)$, where $\omega$ is a primitive element of $GF(4)$ with
$\omega^2+\omega+1=0$, we get the following correspondence between the
Pauli matrices, elements of $GF(4)$, and binary vectors of length two:
\[
\begin{array}{c|c|c}
\text{operator} & GF(4) & \F_2^2\\
\hline
I        & 0        & (00)\\
\sigma_x & 1        & (10)\\
\sigma_y & \omega^2 & (11)\\
\sigma_z & \omega   & (01)
\end{array}
\]
This mapping extends naturally to tensor products of $n$ Pauli matrices
being mapped to vectors of length $n$ over $GF(4)$ or binary vectors
of length $2n$.  We rearrange the latter in such a way that the first
$n$ coordinates correspond to the exponents of the operators
$\sigma_x$ and write the vector as $(a|b)$, i.\,e.,
\begin{equation}\label{eq:binary_rep}
g=\sigma_x^{a_1}\sigma_z^{b_1}\otimes\ldots\otimes\sigma_x^{a_n}\sigma_z^{b_n}
\entspricht(a|b)=(g^X|g^Z).
\end{equation}
Two operators corresponding to the binary vectors $(a|b)$ and $(c|d)$
commute if and only if the symplectic inner product $a\cdot d-b\cdot
c=0$. In terms of the binary representation, the stabilizer
corresponds to a binary code $C$ which is self-orthogonal with respect
to this symplectic inner product, and the normalizer corresponds to
the symplectic dual code $C^*$.  In terms of the correspondence to
vectors over $GF(4)$, the stabilizer and normalizer correspond to an
additive code over $GF(4)$ and its dual with respect to an symplectic
inner product, respectively, which we will also denote by $C$ and
$C^*$.  The term \emph{additive quantum code} refers to this
correspondence.  The minimum distance $d$ of the quantum code is given
as the minimum weight in the set $C^*\setminus C\subset GF(4)^n$ which
is lower bounded by the minimum distance $d^*$ of the additive code
$C^*$.  If $d=d^*$, the code is said to be \emph{pure}, and for $d\ge
d^*$, the code is said to be \emph{pure up to $d^*$}.

Fixing the logical operators $\ol{X}_i$ and $\ol{Z}_j$, there is a
canonical basis for the additive quantum code ${\cal C}$.  The
stabilizer group ${\cal S}$ of the quantum code together with the
logical operators $\ol{Z}_j$ generate an Abelian group of order $2^n$
which corresponds to a self-dual additive code.  The joint
$+1$-eigenspace is one-dimensional, hence there is a unique quantum
state $\ket{\ol{00\ldots0}}\in{\cal C}$ stabilized by all elements of
${\cal S}$.  An orthonormal basis of the code ${\cal C}$ is given by
the states
\begin{equation}\label{eq:canonical_basis_stab}
\ket{\ol{i_1i_2\ldots i_k}}=\ol{X}_1^{i_1}\cdots\ol{X}_k^{i_k}\ket{\ol{00\ldots0}},
\end{equation}
where $(i_1i_2\ldots i_k)\in\F_2^k$.

\subsection{Union stabilizer codes}
The stabilizer group ${\cal S}$ gives rise to an orthogonal
decomposition of the space $(\C^2)^{\otimes n}$ into common
eigenspaces of equal dimension.  The stabilizer code ${\cal C}$ is the
joint $+1$-eigenspace of dimension $2^k$.  In general, the joint
eigenspaces of ${\cal S}$ can be labeled by the eigenvalues of a set
of $n-k$ generators of ${\cal S}$.  Moreover, the $n$-qubit Pauli
group ${\cal P}_n$ operates transitively on the eigenspaces.  Hence
one can identify a set ${\cal T}\subset {\cal P}_n$ of $2^{n-k}$
operators such that
\begin{equation}\label{eq:full_decomposition}
(\C^2)^{\otimes n}=\bigoplus_{t\in{\cal T}} t {\cal C}.
\end{equation}
Note that each of the spaces $t{\cal C}$ is a quantum error-correcting
code with the same parameters as the code ${\cal C}$ and stabilizer
group $t{\cal S}t^{-1}$.  The decomposition
(\ref{eq:full_decomposition}) corresponds to the decomposition of the
$n$-qubit Pauli group ${\cal P}_n$ into cosets with respect to the
normalizer ${\cal N}$ of the code ${\cal C}$ and likewise to the
decomposition of the full vector space $GF(4)^n$ into cosets of the
additive code $C^*$.

The main idea of union stabilizer codes is to find a subset ${\cal
  T}_0$ of the \emph{translations} ${\cal T}$ such that the space
$\bigoplus_{t\in{\cal T}_0} t {\cal C}$ is a good quantum code (see
\cite{GrBe97,GrRo07}).  
\begin{definition}[union stabilizer code]\label{def:unioncode}
Let ${\cal C}_0=[[n,k]]$ be a stabilizer code and let ${\cal
  T}_0=\{t_1,\ldots,t_K\}$ be a subset of the coset representatives of
the normalizer ${\cal N}_0$ of the code ${\cal C}_0$ in ${\cal P}_n$.
Then the \emph{union stabilizer code} is defined as
\[
{\cal C}=\bigoplus_{t\in {\cal T}_0} t{\cal C}_0.
\]
Without loss of generality we assume that ${\cal T}_0$ contains
identity.  The dimension of ${\cal C}$ is $K 2^k$, and we will use the
notation ${\cal C}=((n,K2^k,d))$.
\end{definition}
Similar to (\ref{eq:canonical_basis_stab}) a canonical basis of the union
stabilizer code ${\cal C}$ is given by
\begin{equation}\label{eq:canonical_basis_union}
\ket{\ol{j;i_1i_2\ldots i_k}}=t_j\ol{X}_1^{i_1}\cdots\ol{X}_k^{i_k}\ket{\ol{00\ldots0}},
\end{equation}
where $j=1,\ldots,K$, $(i_1i_2\ldots i_k)\in\F_2^k$, and $\ol{X}_i$
are logical operators of the stabilizer code ${\cal C}_0$.

In order to compute the minimum distance of this code, we first
consider the distance between two spaces $t_1{\cal C}_0$ and $t_2{\cal
  C}_0$.  As for a fixed stabilizer code ${\cal C}_0$ two spaces
$t_1{\cal C}_0$ and $t_2{\cal C}_0$ are either identical or
orthogonal, we can define the distance of them as follows:
\begin{equation}\label{def:distance_pauli}
\dist(t_1{\cal C}_0,t_2{\cal C}_0):=\min\{\wgt(p)\colon p\in{\cal P}_n\mid p t_1{\cal C}_0=t_2{\cal C}_0\}.
\end{equation}
Here $\wgt(p)$ is the number of tensor factors in the $n$-qubit Pauli
operator $p$ that are different from identity.  Clearly,
$\dist(t_1{\cal C}_0,t_2{\cal C}_0)=\dist(t_2^{-1}t_1{\cal C}_0,{\cal
  C}_0)$.  The two spaces are identical if and only if $t_2^{-1}t_1$
is an element of the normalizer group ${\cal N}_0$, or equivalently,
if the cosets $C_0^*+t_1$ and $C_0^*+t_2$ of the additive normalizer
code $C_0^*$ are identical.  (Note that we denote both an $n$-qubit
Pauli operator and the corresponding vector over $GF(4)$ by $t_i$.)
Hence the distance (\ref{def:distance_pauli}) can also be expressed in
terms of the associated vectors over $GF(4)$.
\begin{lemma}\label{lemma:distance}
The distance of the spaces $t_1{\cal C}_0$ and $t_2{\cal C}_0$ equals the
minimum weight in the coset $C_0^*+t_1-t_2$.
\end{lemma}
\begin{IEEEproof}
Direct computation shows
\begin{alignat*}{3}
\dist(t_1{\cal C}_0,t_2{\cal C}_0)
&=\dist(C_0^*+t_1,C_0^*+t_2)\\
&=\dist(C_0^*+(t_1-t_2),C_0^*)\\
&=\min\{\wgt(c+t_1-t_2)\colon c\in C_0^*\}\\
&=\min\{\wgt(v)\colon v \in C_0^*+t_1-t_2\}.
\end{alignat*}
\vskip-1.25\baselineskip
\end{IEEEproof}
While the distance between the cosets $C_0^*+t_j$ is an upper bound on
the minimum distance of the union code ${\cal C}$, the true minimum
distance can be derived from the following code over $GF(4)$.
\begin{definition}[union normalizer code]
With the union stabilizer code ${\cal C}$ we associate the
(in general non-additive) \emph{union normalizer code} given by
\[
C^*=\bigcup_{t\in {\cal T}_0} C_0^*+t=\{c+t_j\colon c \in C_0^*,\, j=1,\ldots,K\},
\]
where $C_0^*$ denotes the additive code associated with the normalizer
${\cal N}_0$ of the stabilizer code ${\cal C}_0$.  We will refer to
both, the vectors $t_i$ and the corresponding unitary operators, as
\emph{translations}.
\end{definition}
\begin{theorem}
The minimum distance of a union stabilizer code with union normalizer
code $C^*$ is given by
\begin{alignat*}{5}
d&=\min\{\wgt(v):v \in(C^*-C^*)\setminus\widetilde{C}_0\}\kern-50pt\\
 &\ge d_{\min}(C^*)\\
&=\min\{\dist(c+t_i,c'+t_{i'})\colon t_i,t_{i'}\in{\cal T}_0,\,&& c,c'\in C^*_0\\
&&& c+t_i\ne c'+t_{i'}\},
\end{alignat*}
where $C^*-C^*:=\{a-b\colon a,b \in C^*\}$ denotes the set of all
differences of vectors in $C^*$, and $\widetilde{C}_0\le C_0$ is the
additive code that corresponds to all elements of the stabilizer group
${\cal S}$ that commute with all $t_j\in{\cal T}_0$.
\end{theorem}
\begin{IEEEproof}
Let $E\in{\cal P}_n$ be an $n$-qubit Pauli error of weight
$0<\wgt(E)<d$. For two canonical basis states $\ket{\psi_a}$ and
$\ket{\psi_b}$ as given in (\ref{eq:canonical_basis_union}) we
consider the inner product
\begin{alignat*}{3}
\bra{\psi_a}E\ket{\psi_b}=&\bra{\ol{j;i_1i_2\ldots i_k}}E\ket{\ol{j';i'_1i'_2\ldots i'_k}}\\
=&\bra{\ol{00\ldots0}} \ol{X}_1^{i_1}\cdots\ol{X}_k^{i_k} t_j
E t_{j'}\ol{X}_1^{i'_1}\cdots\ol{X}_k^{i'_k}\ket{\ol{00\ldots0}}\\
=&\pm\bra{\ol{00\ldots0}} \ol{X}_1^{i_1+i'_1}\cdots\ol{X}_k^{i_k+i'_k}t_j t_{j'} E\ket{\ol{00\ldots0}}.
\end{alignat*}
If $E\in{\cal S}$ commutes with all $t_j\in{\cal T}_0$, then
$\bra{\psi_a}E\ket{\psi_b}=\delta_{ab}$. Otherwise, $E\notin C^*-C^*$
since $0<\wgt(E)<d$, and hence the inner product vanishes.
\end{IEEEproof}

\section{The Binary Goethals and Preparata Codes}
In this section we recall some properties of the binary Goethals codes
\cite{Goe74} and the Preparata codes \cite{Pre68}.  It has been shown
that variations of these codes have a simple description as
$\Z_4$-linear codes \cite{HKC94}, but in our context the description in
terms of cosets of linear binary codes is used.

In the following $m$ is an even integer ($m\ge 4$) and
$n=2^{m-1}-1$. Let$\alpha$ be a primitive element of the finite field
$GF(2^{m-1})$. By $\mu_i(z)$ we denote the minimal polynomial of
$\alpha^i$ over $GF(2)$, i.\,e., the polynomial with roots $\alpha^j$
for $j=i2^k$.  The idempotent $\theta_i(z)$ is the unique polynomial
satisfying
\[
\theta_i(\alpha^i)=1\quad\text{and $\quad\theta_i(\alpha^j)=0$ for $j\ne i2^k$.}
\]
Codewords of a cyclic code can be represented by  polynomials $f(z)$,
and we use $(f(z);f(1))$ to denote the codeword of the extended cyclic
code obtained by adding an overall parity check.  Similar, we use
$(f(z);f(1);g(z);g(1))$ to denote the juxtaposition of codewords of two
extended cyclic codes.

\begin{definition}[Goethals code \cite{Goe74}]\label{def:Goethals}
The Goethals code ${\cal G}(m)$ of length $2^m$ is the union of
$2^{m-1}$ cosets of the linear binary code $C_{\cal
  G}=[2^m,2^m-4m+2,8]$. The code $C_{\cal G}$ is obtained via the
$|u|u+v|$ construction applied to the extended cyclic codes
$\ol{C_1}$ and $\ol{C_2}$. The cyclic code $C_1$ is a
single-error correcting code with generator polynomial $\mu_1(z)$, and
$C_2$ is generated by $\mu_1(z)\mu_r(z)\mu_s(z)$ where $r=1+2^{m/2-2}$
and $s=1+2^{m/2-1}$.  The non-zero coset representatives are given by
$(z^i;1;z^i\theta_1(z);0)$ for $i=1,\ldots,n-1$.
\end{definition}

An alternative description of Goethals codes has been given in
\cite{BvLW83}. The codewords are described by pairs $(X,Y)$ of subsets
of $GF(2^{m-1})$. The corresponding codeword is given by the
juxtaposition of the characteristic functions $\chi_X$ and $\chi_Y$ of
the two set $X$ and $Y$, i.\,e.
\[
(X,Y) \entspricht
(1_X(\alpha^i);1_X(0);1_Y(\alpha^i);1_Y(0)),
\]
where $1_X(\alpha^i)$ is a short-hand for the vector
\[
1_X(\alpha^i)=(1_X(\alpha^0),1_X(\alpha^1),\ldots,1_X(\alpha^{n-1}))
\]
and
\[
1_S(x)=\begin{cases}
1 & \text{if $x \in S$},\\
0 & \text{if $x \notin S$}.
\end{cases}
\]
The non-zero elements of $X$ and $Y$ give rise to the polynomials
$f_X(z)$ and $f_Y(z)$ given by
\begin{equation}\label{eq:f_S}
f_S(z)=\sum_{i=0}^{n-1} 1_S(\alpha^i) z^i.
\end{equation}

\begin{definition}[Goethals code \cite{BvLW83}]\label{def:Goethals2}
The Goethals code ${\cal G}(m)$ of length $2^m$ consists of the
codewords described by all pairs $(X,Y)$ satisfying:
{\def\theenumi{\alph{enumi}}
\begin{enumerate}
\item $|X|$ is even, $|Y|$ is even,
\item $\displaystyle\sum_{x\in X} x=\sum_{y\in Y} y$,
\item $\displaystyle\sum_{x\in X} x^r+\left(\sum_{x\in X} x\right)^r=\sum_{y\in Y} y^r$,
\item $\displaystyle\sum_{x\in X} x^s+\left(\sum_{x\in X} x\right)^s=\sum_{y\in Y} y^s$.
\end{enumerate}
}%
\end{definition}

In order to relate the two definitions, we distinguish three cases.
\begin{enumerate}
\item $X=Y$: Conditions c) and d) imply that $\sum_{x\in  X}x=0$. This
  is true for all codewords of the cyclic code generated by
  $\mu_1(z)$.  Adding an overall parity check implies Condition
  a).
\item $X=\emptyset$: The left hand side of Conditions b), c), and d)
  vanishes, so the solutions for $Y$ correspond to an extended cyclic
  code with generator polynomial $\mu_1(z)\mu_r(z)\mu_s(z)$.
\item $X=\{0,x=\alpha^i\}$: From (\ref{eq:f_S}) it follows that
  $f_Y(\alpha)=\sum_{y\in Y} y$.  So Condition b) holds for the set
  $Y$ corresponding to $f_Y(z)=z^i\theta_1(z)$. The left hand side of
  Conditions c) and d) vanishes, so the solutions for $Y$ are elements
  of the extended cyclic code with generator polynomial
  $\mu_r(z)\mu_s(z)$.  As neither $r$ nor $s$ is a power of two, the
  polynomial $\theta_1(z)$ and hence $f_Y(z)=z^i\theta_1(z)$ vanishes
  for $\alpha^r$ and $\alpha^s$, i.\,e., Conditions c) and d) hold.
\end{enumerate}
Finally, all codewords of the Goethals code as given in
Definition~\ref{def:Goethals} are the juxtaposition of two binary
vectors of even weight, i.\,e., Condition a) holds.  Hence any
codeword given by Definition~\ref{def:Goethals} fulfills the
conditions of Definition~\ref{def:Goethals2}. The equivalence of the
definitions follows from the fact that the codes have equal size.

Next we consider the definition of Preparata codes similar to
Definition~\ref{def:Goethals2} given in \cite{BvLW83}.
\begin{definition}[Preparata code \cite{BvLW83}]\label{def:Preparata2}
The extended Preparata code ${\cal P}(m)$ of length $2^m$ and
parameter $\sigma$ consists of the codewords described by all pairs
$(X,Y)$ satisfying: {\def\theenumi{\alph{enumi}}
\begin{enumerate}
\item $|X|$ is even, $|Y|$ is even,
\item $\displaystyle\sum_{x\in X} x=\sum_{y\in Y} y$,
\item $\displaystyle\sum_{x\in X} x^{\sigma+1}+\left(\sum_{x\in X} x\right)^{\sigma+1}=\sum_{y\in Y} y^{\sigma+1}$,
\end{enumerate}
}%
Here $\sigma$ is a power of two and $\gcd(\sigma\pm1,n)=1$.
\end{definition}
For $\sigma=2^{m/2-1}$ and $n=2^{m-1}-1$ we compute
\[
\left(2^{m-1}-1\right)-\left(2^{m/2-1}\pm1\right)\left(2^{m/2}\mp2\right)=1,
\]
showing that $\gcd(\sigma\pm1,n)=1$.  Hence for this particular choice
of $\sigma$, the Preparata code of Definition~\ref{def:Preparata2}
contains the Goethals code. What is even more, we can describe the
Preparata code similar to Definition~\ref{def:Goethals} as the union
of cosets of the linear binary code $C_{\cal P}$ which contains the
linear binary code $C_{\cal G}$.
\begin{definition}\label{def:Preparata}
The extended Preparata code ${\cal P}(m)$ of length $2^m$ is the union
of $2^{m-1}$ cosets of the linear binary code $C_{\cal
  P}=[2^m,2^m-3m+1,6]$.  The code $C_{\cal P}$ is obtained via the
$|u|u+v|$ construction applied to the extended cyclic codes
$\ol{C_1}$ and $\ol{C_3}$. The cyclic code $C_1$ is a
single-error correcting code with generator polynomial $\mu_1(z)$, and
$C_3$ is generated by $\mu_1(z)\mu_s(z)$ where $s=1+2^{m/2-1}$. The
non-zero coset representatives are given by
$(z^i;1;z^i\theta_1(z);0)$.
\end{definition}
Comparing Definitions~\ref{def:Goethals} and \ref{def:Preparata} we
see that we can use the very same coset representatives to construct
the Goethals and the Preparata code as union of cosets of the linear
binary codes $C_{\cal G}$ and $C_{\cal P}$, respectively. Moreover,
all codes lie between codes that are equivalent to the Reed-Muller
codes $RM(m-3,m)$ and $RM(m-2,m)=[2^m,2^m-m-1,4]$ (see
\cite{Her90}). This is illustrated by the following diagram:
\[
\begin{picture}(130,140)(-40,-30)
\put(10,-20){\makebox(0,0){${\rm RM}(m-3,m)$}}
\put(10,-13){\line(0,1){16}}
\put(10,10){\makebox(0,0){\llap{$[2^m,2^m-4m+2,8]={}$}$C_{\cal G}$}}
\put(10,17){\line(0,1){26}}
\put(10,50){\makebox(0,0){\llap{$[2^m,2^m-3m+1,6]={}$}$C_{\cal P}$}}
\put(20,13){\line(4,1){25}}
\put(20,53){\line(4,1){25}}
\put(50,20){\makebox(0,0)[l]{${\cal G}(m)=\bigcup_i C_{\cal G}+t_i$}}
\put(55,27){\line(0,1){26}}
\put(50,60){\makebox(0,0)[l]{${\cal P}(m)=\bigcup_i C_{\cal P}+t_i$}}
\put(55,67){\line(0,1){16}}
\put(50,90){\makebox(0,0)[l]{\llap{$[2^m,2^m-m-1,4]={}$}${\rm RM}(m-2,m)$}}
\end{picture}
\]
The components of the codes are summarized as follows:
\begin{description}
\item[$C_1$:] cyclic code generated by $\mu_1(z)$
\item[$C_3$:] cyclic code generated by $\mu_1(z)\mu_s(z)$
\item[$C_2$:] cyclic code generated by $\mu_1(z)\mu_r(z)\mu_s(z)$\\
   $r=1+2^{m/2-2}$, $s=1+2^{m/2-1}$
\item[$C_{\cal G}$:] $|u|u+v|$ construction applied to the extended cyclic
  codes $\ol{C_1}$ and $\ol{C_2}$
\item[$C_{\cal P}$:] $|u|u+v|$ construction applied to the extended cyclic
  codes $\ol{C_1}$ and $\ol{C_3}$
\item[$t_i$:] $n+1$ coset representatives with
\[
t_i=\begin{cases}
(z^i;1;z^i \theta_1(z);0) & \text{for $i=0,\ldots,n-1$},\\
(0,\ldots,0) & \text{for $i=n$}.
\end{cases}
\]
\end{description}

\section{The Quantum Goethals-Preparata Codes}
Before presenting the new family of non-additive quantum codes, we
recall Steane's construction to enlarge the dimension of CSS codes.
\begin{theorem}[see \cite{Stea99b}]
Let $C=[n,k,d]$ and $C'=[n,k'>k+1,d']$ be linear binary codes with
$C^\bot\le C< C'$.  Then there exists an additive quantum code ${\cal
  C}=[[n,k+k'-n,\ge\min(d,3d'/2)]]$.  Given a generator matrix $G$ of
the code $C$ and a generator matrix $D$ of the complement of $C$ in
$C'$, the normalizer of the code ${\cal C}$ is generated by
\[
\left(
\begin{array}{c|c}
G & 0\\
0 & G\\
D & A D
\end{array}
\right),
\]
where $A$ is a fixed-point free linear transformation.
\end{theorem}
As the code $C_{\cal G}$ contains a code that is isomorphic to the
Reed-Muller code $RM(m-3,m)$ it follows that $C_{\cal G}^\bot\le
C_{\cal G}$. Hence we can apply Steane's construction \cite{Stea99b}
to the chain $C_{\cal G}^\bot\le C_{\cal G}< C_{\cal P}$ of linear
binary codes and obtain an additive quantum code with parameters
${\cal C}_0=[[2^m,2^m-7m+3,8]]$.  In a second step we use the
$K=2^{m-1}$ coset representatives $t_i$ of the decomposition of both
the Goethals and the Preparata code.  This yields a non-additive code
with dimension $K^22^{2^m-7m+3}=2^\ell$ where $\ell=2^m-5m+1$.

\begin{figure}[thb]
\[
\begin{array}{c}
\left(\def\arraystretch{1.3}
\begin{array}{c|c}
G&0\\
0&G\\
D&AD\\
\hline
\end{array}
\right)\\
\left\{\def\arraystretch{1.3}
\begin{array}{c|c}
t_1&t_1\\[-1.2ex]
\vdots&\vdots\\[-1.0ex]
t_1&t_K\\
\vdots&\vdots\\
t_K&t_1\\[-1.2ex]
\vdots&\vdots\\[-1.0ex]
t_K&t_K
\end{array}
\right\}
\end{array}
\]
\caption{Structure of the non-additive union normalizer code of the
  quantum Goethals-Preparata codes.\label{fig:union_CSS_generators}}
\end{figure}

\begin{theorem}
Let ${\cal C}_0=[[2^m,2^m-7m+3,8]]$ be the additive quantum code
obtained from the chain of linear binary codes $C_{\cal G}^\bot \le
C_{\cal G}\le C_{\cal P}$ using Steane's enlargement construction.
Furthermore, let ${\cal T}_0=\{(t_i|t_j)\colon
i,j=0,\ldots,2^{m-1}-1\}$ where $t_i$ are the coset representatives
used to obtain the Goethals and Preparata code.  Then the {\em quantum
  Goethals-Preparata code} is a union stabilizer code given by ${\cal
  C}_0$ and ${\cal T}_0$.  The minimum distance of the quantum
Goethals-Preparata code is eight.
\end{theorem}
\begin{IEEEproof}
Let $G$ denote a generator matrix of the code $C_{\cal G}$ and let $D$
be such that $\left(\begin{smallmatrix}G\\D\end{smallmatrix}\right)$
generates $C_{\cal P}$. The structure of the non-additive
union-normalizer code of the quantum Goethals-Preparata codes is
illustrated in Fig. \ref{fig:union_CSS_generators}.  A generator
matrix of the normalizer of the additive quantum code ${\cal C}_0$ is
given above the horizontal line, while the set of translations is
listed below the horizontal line.  Every codeword of the non-additive
union normalizer code is of the form
\[
g=(g^X|g^Z)=(c_1+v+t_i|c_2+w+t_j),
\]
where $c_1,c_2\in C_{\cal G}=[2^m,2^m-4m+2,8]$ and $v,w\in C_{\cal  P}/C_{\cal G}$. 
For $g,g'\in C^*,g\ne g'$ we compute
\begin{alignat*}{5}
\dist(g,g')=&\dist(&&(c_1+v+t_i|c_2+w+t_j),\\
            &      &&(c'_1+v'+t'_i|c'_2+w'+t'_j))\\
=&\wgt(&&(c''_1+v''+t_i-t'_i|c''_2+w''+t_j-t'_j)),
\end{alignat*}
where $c''_1=c_1-c'_1$ and $c''_1=c_1-c'_1$ are codewords of $C_{\cal G}$,
and $v''=v-v'$, $w''=w-w'$ are codewords of $C_{\cal P}/C_{\cal G}$. In
general, the weight of $g=(g^X|g^Z)$ is given by
\[
\wgt((g^X|g^Z))=\frac{1}{2}(\wgt(g^X)+\wgt(g^Z)+\wgt(g^X+g^Z)).
\]
Hence we get
\begin{subequations}
\begin{alignat}{3}
\dist&(g,g')=\nonumber\\
 &\frac{1}{2}\wgt(c''_1+v''+t_i-t'_i)\label{eq:wgt1}\\
+&\frac{1}{2}\wgt(c''_2+w''+t_j-t'_j)\label{eq:wgt2}\\
+&\frac{1}{2}\wgt(c''_1+c''_2+v''+w''+t_i-t'_i+t_j-t'_j).\label{eq:wgt3}
\end{alignat}
\end{subequations}
By Steane's construction the vectors $v''$ and $w''$ are either both
zero, or both are non-zero and they are different.  For $v''=w''=0$,
we can assume without loss of generality that the vectors in
(\ref{eq:wgt1}) and (\ref{eq:wgt2}) are both non-zero.  The weight of
these vectors equals the distance between two codewords of the
Goethals code, so it is at least 8.  For $v''\ne 0\ne w''$ the terms
(\ref{eq:wgt1}) and (\ref{eq:wgt2}) equal the distance of two
codewords of the Preparata code, so they are lower bounded by 6.  We
will show that for $v''\ne w''$, the vector in (\ref{eq:wgt3}) is a
non-zero codeword of the linear code isomorphic to the Reed-Muller
code $RM(m-2,m)$, hence its weight is at least 4. For this, consider
the vectors $a=(a_1;a_2)=c''_1+c''_2+v''+w''\ne 0$ and
$b=(b_1;b_2)=t_i-t'_i+t_j-t'_j$.  The coset representatives are of the
form $t_i=(z^i;1;z^i\theta_1(z);0)$, so the second half $b_2$ of $b$
is a codeword of the extended cyclic code generated by $\theta_1(z)$,
while $a_2$ is a codeword of the extended cyclic code generated by
$\mu_1(z)$. The intersection of the two codes is trivial, so $a_2=b_2$
only if $a_2=b_2=0$. Then $\wgt(b)\le 4$ while $\wgt(a)\ge 6$ since
$0\ne a\in C_{\cal P}$.  Hence $a\ne b$.
\end{IEEEproof}
To our best knowledge, the best additive quantum code with the same
length and minimum distance has dimension $2^m-5m-2$.  Codes with
these parameters can, e.\,g., be obtained by applying Steane's
construction to extended primitive BCH codes $[2^m,2^m-2m-1,8]$ and
$[2^m,2^m-3m-1,6]$ (see \cite{Stea99b}). In the following table we
give the parameters of the first codes in these families.
Additionally, we give the parameters of the non-additive quantum codes
derived from Goethals codes in \cite{GrRo07}.
\[\def\arraystretch{1.3}
\begin{array}{c|c|c}
\text{Goethals}& \text{enlarged BCH}& \text{Goethals-Preparata}\\
\hline
((  64, 2^{ 30}, 8))& [[   64,  32, 8]] & ((   64, 2^{35},  8)) \\{}
(( 256, 2^{210}, 8))& [[  256, 214, 8]] & ((  256, 2^{217}, 8)) \\{}
((1024, 2^{966}, 8))& [[ 1024, 972, 8]] & (( 1024, 2^{975}, 8)) 
\end{array}
\]

\section{Conclusions}
We have constructed some new non-additive quantum codes from nested
non-linear binary codes which can be decomposed into cosets of linear
codes which contain their dual.  It is interesting to find more good
non-linear binary or quaternary codes with this property.

Recently, Ling and Sol{\'e} have constructed some non-additive quantum
codes from $\Z_4$-linear codes using a CSS-like construction
\cite{LiSo07}.  So far it is not clear whether the non-additive codes
presented here can also be put into the framework of
$\Z_4$-linear codes.

\section*{Acknowledgments}
We acknowledge fruitful discussions with Vaneet Aggarwal and Robert
Calderbank.  Markus Grassl would like to thank NEC Labs., Princeton
for the hospitality during his visit as well as Tero Laihonen, Kalle
Ranto, and Sanna Ranto for discussions on variations of Goethals
codes.  This work was partially supported by the FWF (project P17838).

\end{document}